\documentclass[authoryear,12pt,1p]{elsarticle}

\newcommand{\E}{\ensuremath{\mathbb{E}}}
\newcommand{\VAR}{\ensuremath{\mathbb{V}\text{ar}}}

\usepackage{float}

\usepackage{ifthen}
\newcommand{\mynote}[1]{%
  \ifthenelse{\boolean{notesBoolean}}{
    \begingroup
    \hbadness 20000
    \marginpar{\tiny\textsf{#1}}
    \endgroup
  }{}
}

\usepackage{amsmath,amssymb,amsthm}
\usepackage{bm}
\usepackage{natbib}

\DeclareMathOperator{\logit}{logit}
\DeclareMathOperator{\id}{id}

\newcommand{\Tinf}{\ensuremath{\mathfrak{T}_\infty}}
\newcommand{\Tto}{\ensuremath{\mathfrak{T}_2}}

\newcommand{\var}{\VAR}
\newcommand{\pr}{\Prob}
\newcommand{\one}{\bm{1}}
\newcommand{\R}{\ensuremath{\mathbb{R}}}

\newcommand{\abs}[1]{\ensuremath{\left\vert#1\right\vert}}

\renewcommand{\pr}{\ensuremath{\mathbb{P}}}

\usepackage{Sweave}
\newcommand{\proglang}[1]{\texttt{#1}}
\newcommand{\pkg}[1]{\texttt{#1}}
\newcommand{\code}[1]{\texttt{#1}}
\usepackage{url}

\renewenvironment{Schunk}{
\small
}
{
\normalsize
}

\begin{document}
\begin{frontmatter}
  \title{Model Diagnostics Based on Cumulative Residuals: The R-package \pkg{gof}}
  \author[biostat]{Klaus K\"{a}hler Holst}
  \ead{k.k.holst@biostat.ku.dk}
  \address[biostat]{University of Copenhagen, Department of
    Biostatistics}
  \begin{abstract}
    The generalized linear model is widely used in all areas of
applied statistics and while correct asymptotic inference can be
achieved under misspecification of the distributional assumptions, a
correctly specified mean structure is crucial to obtain interpretable
results. Usually the linearity and functional form of predictors are
checked by inspecting various scatterplots of the residuals, however,
the subjective task of judging these can be 
challenging. In this paper we present an implementation of model
diagnostics for the generalized linear model as well as
structural equation models, based on aggregates of the residuals
where the asymptotic behavior under the null is imitated by
simulations. A procedure for checking the proportional hazard
assumption in the Cox regression is also implemented.


  \end{abstract}
\end{frontmatter}



\
\section{Introduction}
The generalized linear model is one of the most widely used classes of
statistical models, however, the standard methods of inference relies
on distributional and linearity assumptions. The importance of this is
sometimes underestimated, to some extent because few tools are
available for checking all the aspects of the model. While the
distributional assumptions can be relaxed, i.e., by using a sandwich
estimator as implemented in the \pkg{sandwich} package
\citep{Zeileis:2006nx}, careful attention should be paid
to the validity of the specified mean structure. A typical model check
involves assessment of various residual plots.  As the true variance
of individual residuals are unknown it can be difficult to
decide whether a residual plot indicates a reasonable specification of
the mean or not.  In a paper by \cite{MR1137124} it was
proposed instead to look at certain aggregates of the residuals, such
as the cumulative sum over predicted values or covariates. The key
result here is, that the asymptotic distribution of such aggregates
can be determined under the hypothesis that the model is
correctly specified.

\vspace*{0.5cm}

The \proglang{R} environment is one of the most widely used statistics platforms
but lacks objective diagnostics tools for many regression models,
and in particular methods based on aggregates of residuals, thus
motivating the creation of the \pkg{gof}-package described in the
following sections.

\section{Implementation}

The \pkg{gof} package implements diagnostics of the linearity
assumptions for the generalized linear model and linear structural
equation models. Further similar methods are available for checking
the proportional hazards assumption of the Cox regression model for
right censored data. The following section describes the theoretical
details behind the implementation.

\subsection{Generalized linear model}

The case of generalized linear models was first examined by
\cite{MR1137124}. Let $Y$ be the response variable with a distribution
from a (natural) exponential family:
\begin{align}
  f(Y=y_i \mid \theta_i,\phi) = \exp\left\{\tfrac{\theta_i
      y-b(\theta_i)}{a(\phi)}+c(y_i,\phi)\right\},
\end{align}
parameterized by $\theta$ (and the dispersion parameter $\phi$) and the
known functions $a, b$ and $c$.  Direct calculations reveals that the
\begin{align}
  \E Y_i = b'(\theta_i), \quad \var(Y_i) = a(\phi)b''(\theta_i).
\end{align}
The mean $\E{Y_i} = \mu(\theta_i)$ is
related to some covariates, $\bm{x}_i$, through a link-function \citep{MR727836}, $g$, 
\begin{align} 
  g\left\{\mu(\theta_i)\right\} = \bm{\beta}^T\bm{x}_i,
\end{align}
i.e., $\theta_i = \theta_i(\bm{\beta})$.  Typically, the canonical link
is chosen such that $g\circ \mu =\id$, with the most common
regression models being the general linear model, logistic
regression and Poisson regression
\begin{center}
  \begin{tabular}{llll}
    family & canonical link & $a(\phi)$ & $b'(\theta)$\\ \hline
    Normal & identity & $\phi$ & $\theta$ \\
    Binomial & logit & 1 & $1/(1+\exp(-\theta))$\\
    Poisson & log & 1 & $\exp(\theta)$ \\  \hline
  \end{tabular}    
\end{center}

Given $n$ observations $(y_i,x_{1i},\ldots,x_{pi})_{i=1,\ldots,n}$ the
maximum likelihood estimate $\widehat{\bm{\beta}}\in\R^{p}$ is
obtained by solving the set of score equations:
\begin{align}
  U(\bm{\beta}) = \sum_{i=1}^n h(\bm{\beta}^T\bm{x}_i)\bm{x}_i\left\{y_i-g^{-1}(\bm{\beta}^T\bm{x}_i)\right\},
\end{align}
with $h = \partial\{(g\circ \mu)^{-1}\}$.
We define the (raw) residuals $e_i = y_i -
g^{-1}(\widehat{\bm{\beta}}^T\bm{x}_i)$, $i=1,\ldots,n$.
Our interest is the cumulative sum of the residuals over the $j$th
covariate \citep{MR1137124, lin02:_model_check_techn_based_cumul_resid}:
\begin{align}\label{eq:Wj}
  W_j(x) = n^{-1/2}\sum_{i=1}^n \one_{\{x_{ji}\leq x\}}e_i.
\end{align}
In contrast to the distribution of individual residuals, we can
determine the distribution (under the null) of this aggregate. For
known parameters the asymptotics can be derived as a Brownian bridge
\citep{shorack86:_empir_proces_applic_statis}, however, we need to take uncertainty in
estimation of $\widehat{\bm{\beta}}$ into account.
Under certain regularity conditions, a Taylor expansion around the
true parameter value, $\bm{\beta}_0$, gives us
\begin{align}
  W_j(x) = W_j(x\mid\widehat{\bm{\beta}}) = W_j(x\mid \bm{\beta}_0) +
  \left.\frac{\partial}{\partial\bm{\beta}}W_j(x\mid
    \bm{\beta})\right|_{\bm{\beta}=\bm{\beta_0}}(\widehat{\bm{\beta}}-\bm{\beta}_0)
  + o_p(1).
\end{align}
Let $\mathcal{I}(\widehat{\bm{\beta}}) = \E(-\nabla
U(\widehat{\bm{\beta}}))$ denote the Fisher information. Now
$(\widehat{\bm{\beta}}-\bm{\beta}_0)$ is asymptotically normally
distributed and asymptotically equivalent with
$\mathcal{I}(\widehat{\bm{\beta}})^{-1}U(\widehat{\bm{\beta}})$:
\begin{align}
  (\widehat{\bm{\beta}}-\bm{\beta}_0) =
  \mathcal{I}(\widehat{\bm{\beta}})^{-1}U(\widehat{\bm{\beta}})
  + o_p(1).
\end{align}
It then follows that the process
\begin{align}\label{eq:What}
  \widehat{W}_j(x) = n^{-1/2}\sum_{i=1}^n\left[\one_{\{x_{ji}<x\}} +
    \bm{\eta}_j(x\mid\widehat{\bm{\beta}})\mathcal{I}^{-1}(\widehat{\bm{\beta}})\bm{x}_ih(\widehat{\bm{\beta}}^T\bm{x}_i)
  \right]e_i G_i
\end{align}
with i.i.d. $G_1,\ldots,G_n\sim\mathcal{N}(0,1)$, $i=1,\ldots,n$, and
\begin{align}
  \bm{\eta}_j(x\mid\bm{\beta}) =
  -\sum_{i=1}^n\one_{\{x_{ji}\leq x\}}\frac{\partial g^{-1}(\bm{\beta}^T\bm{x}_i)}{\partial\bm{\beta}},
\end{align}
(see Table \ref{tab:linkfun}) converges weakly to the same limiting
distribution as the observed process (\ref{eq:Wj}) \citep{lin02:_model_check_techn_based_cumul_resid}.
\begin{table}[htbp!]
  \centering
  \begin{tabular}{c|c|c}
    $g(x)$ & $g^{-1}(z)$ & $\partial(g^{-1})(z)$  \\ \hline
    $x$ & $z$ & $1$ \\
    $\logit(x)$  &  $1/(1+\exp(-z))$ & $\exp(-z)/[1+\exp(-z)]^2$ $ $ \\
    $\log(x)$ & $\exp(z)$ & $\exp(z)$ \\ \hline
  \end{tabular}  
  \caption{Some link functions and their inverse.}
  \label{tab:linkfun}
\end{table}

To test the functional form of the $j$th covariate we look at a
Kolmogorov-Smirnov (KS) type supremum statistic:
\begin{align}
  \Tinf^{(j)}\colon W_j\mapsto \sup_x \abs{W_j(x)}.
\end{align}
Alternatively tests can be based on the Cramer-von-Mises (CvM)
functional:
\begin{align}
  \Tto^{(j)}\colon W_j\mapsto \int\abs{W_j(x)}^2\,dx.
\end{align}
A large number of realizations of $\widehat{W}_j$ is generated. The
supremum statistic is calculated for each realization and the p-value
is estimated from the empirical distribution of these statistics.
The residuals can also be cumulated after the predicted values
\citep{lin02:_model_check_techn_based_cumul_resid}
\begin{align}\label{eq:Wy}
  W_{\widehat{\bm{y}}}(t) = n^{-1/2}\sum_{i=1}^n \one_{\{g^{-1}(\widehat{\bm{\beta}}^T\bm{x}_{i})\leq t\}}e_i,
\end{align}
which leads to a test of misspecified link function.

\subsection{Structural equation models}
The linear structural equation models covers a broad range of models
including the general linear model, path analysis and various latent
variable models.  Diagnostics based on cumulative residuals was
examined in this case by \cite{BNSanchez2009} building on the work of
\cite{ZhiyingPan2005} on Generalized Linear Mixed Models (GLMM) sharing
many of the aspects of structural equation models. The basic idea and
proof of weak convergence is very similar to the case of GLM.

A structural equation model is typically divided into two separate parts.
For the $i$th individual we have a \emph{measurement part} describing
the multivariate outcome $\bm{Y}_i$:
\begin{align}
  \bm{Y}_i = \bm{\nu} + \bm{\Lambda}\bm{\eta}_i + \bm{K}\bm{X}_i +
  \bm{\epsilon}_i,
\end{align}
where $\bm{\eta}_i$ are the latent variables and $\bm{X}_i$ are covariates,
and a \emph{structural part} describing the latent variables:
\begin{align}
  \bm{\eta}_i = \bm{\alpha} + \bm{B}\bm{\eta}_i +
  \bm{\Gamma}\bm{X}_i + \bm{\zeta}_i
\end{align}  
where $\bm{\nu}\in\R^{p}$, $\bm{\Lambda}\in\R^{p\times l}$,
$\bm{K}\in\R^{p\times q}$, and
$\bm{\epsilon}_i\sim\mathcal{N}_p(0,\bm{\Sigma_{\epsilon}})$. And $\bm{\alpha}\in\R^l$,
$\bm{B}\in\R^{l\times l}$, $\bm{\Gamma}\in\R^{l\times q}$,
and $\bm{\zeta}_i\sim\mathcal{N}(0,\bm{\Psi})$.
Hence, the model is parameterized by some $\bm{\theta}$ defining
$(\bm{\nu},\bm{\alpha},\bm{\Lambda},\bm{K},\bm{B},\bm{\Gamma},\bm{\Sigma_{\epsilon}},$
$\bm{\Psi})$
with some restrictions to guarantee identification.  The conditional
moments of $\bm{Y}_i$ given $\bm{X}_i$, are
\begin{align}
  \begin{split}
    \bm{\mu}_i = \E_{\bm{\theta}}(\bm{Y}_i\mid\bm{X}_i) &= \bm{\nu} +
    \bm{\Lambda}(\one-\bm{B})^{-1}\bm{\alpha} \\ &\qquad +
    \left[\bm{\Lambda}(\one-\bm{B})^{-1}\bm{\Gamma} +
      \bm{K}\right]\bm{Y_i},
  \end{split}
\end{align}
\begin{align}
  \bm{\Sigma} = \var_{\bm{\theta}}(\bm{Y}_i\mid \bm{X}_i) &= 
  \bm{\Lambda}(\one-\bm{B})^{-1}\bm{\Psi}(\one-\bm{B})^{-1}{}^T\bm{\Lambda}^T,
\end{align}
and inference on $\bm{\theta}$ is usually obtained by MLE \cite{MR996025}. 

The residuals can be predicted as the conditional mean given the
endogenous variables and covariates. Hence,
\begin{align}
  \widehat{\epsilon}_{ik} = \E(\epsilon_{ik}\mid \bm{Y}_i,\bm{X}_i) =
  \bm{\pi}_k^p\bm{\Sigma_{\epsilon}}\bm{\Sigma}^{-1}(\bm{Y}_i-\bm{\mu}_i)
\end{align}
\begin{align}
  \widehat{\zeta}_{ig} = \E(\zeta_{ig}\mid \bm{Y}_i,\bm{X}_i) =
  \bm{\pi}_g^l\bm{\Psi}(\one-\bm{B})^{-1}{}^T\bm{\Lambda}^T\bm{\Sigma}^{-1}(\bm{Y}_i-\bm{\mu}_i)
\end{align}
where $\bm{\pi}_r^s\colon \R^s\to\R$ is the projection onto coordinate $s$.
Different local aspects of the structural equation model can now be
assessed by examining the cumulative residual processes of either
$\widehat{\epsilon}_{ik}$ and $\widehat{\zeta}_{ig}$. 

\emph{Misspecified covariate effect on the $g$th latent variable} is
checked by summing $\widehat{\zeta}_{ig}$ with respect to the $j$th covariate, $(X_{ij})$:
\begin{align}
  W^l_{\bm{X}_j}(x) = n^{-1/2}\sum_{i=1}^n \one_{\{X_{ij}\leq x\}}\widehat{\zeta}_{ig}
\end{align}
and as for the GLM we can imitate the behavior of this process under
the null of no misspecification by simulation. \emph{Misspecified link
  between $g$th latent variable and its predictors} is checked by
summing $\widehat{\zeta}_{ig}$ with respect to $\E(\eta_{ig} \mid
\bm{X_i}) =
\bm{\pi}_g^l(\one-\bm{B})^{-1}(\bm{\alpha}+\bm{\gamma}\bm{X}_i)$.  To
examine \emph{departures from the specified association between an
  endogenous variable} and one of its predictors, we can look at the
cumulative process defined by summing $\widehat{\epsilon}_{ik}$ with
respect to $X_j$ or $\E(\eta_{ig}\mid \bm{X}_i)$. This can also be
used to diagnose for so-called \emph{item bias} (conditional
dependence between a covariate and endogenous variable given latent
variables).  Finally, \emph{misspecified link between an endogenous
  variable and its linear predictors} is checked by summing
$\widehat{\epsilon}_{ik}$ with respect to $\E(Y_{ik} \mid \bm{X}_i)$.
  
\subsection{Cox's proportional hazard model}
The idea of looking at aggregates of residuals can also be applied as
a tool for diagnosing the proportional hazards assumptions used in
many survival analyses. We will assume that we have triplet
observations $(N_i(t),$ $Y_i(t),$ $X_i(t)),$ $i=1,\ldots,n$ of a counting
process, at-risk process and covariate process in the compact
time-interval $[0,\tau]$. Using the notation of stochastic integrals
we let the Martingale decomposition of the counting process be given by
\begin{align}
  dN_i(t) = \lambda_i(t)\,dt + dM_i(t).
\end{align}
Cox's proportional hazard model assumes intensity takes the form
\begin{align}
  \lambda_i(t) = Y_i(t)\lambda_0(t)\exp(X_i^T(t)\beta),
\end{align}
where $X$ is $p$-dimensional covariates. We denote the cumulative
baseline hazard
\begin{align}
  \Lambda_0(t) = \int_0^t\lambda_0(s)\,ds.
\end{align}
As the model contains a non-parametric term, $\lambda_0$, inference
will be based on the partial likelihood \citep{cox72}
\begin{align}
  L(\beta) = \prod_{i=1}^n \prod_t
  \left(\frac{\exp(X_i^T(t)\beta)}{S_0(t,\beta)}\right)^{\Delta  N_i(t)}
\end{align}
where
\begin{align}
  S_0(t,\beta) = \sum_{i}Y_i(t)\exp(X_i^T(t)\beta).
\end{align}
with the first and second partial derivatives 
\begin{align}
  S_1(t,\beta) = \sum_{i}Y_i(t)\exp(X_i^T(t)\beta)X_i(t), \\
  S_2(t,\beta) = \sum_{i}Y_i(t)\exp(X_i^T(t)\beta)X_i(t)^{\otimes 2},
\end{align}
and let $E(t,\beta) = S_1/S_0(t,\beta)$.
The score equation then becomes
\begin{align}
  U(\beta) = \sum_{i=1}^n \int_0^\tau \left[X_i(t)-E(t,\beta)\right]\,dN_i(t).
\end{align}
The Nelson-Aalen estimator of the cumulative intensity is
\begin{align}
  \widehat{\Lambda}_0(t) = \int_0^t \frac{1}{S_0(s,\beta)}\,dN_{.}(s),
\end{align}
where $N_{.} = \sum_iN_i$.
Define
\begin{align}
  I(t,\beta) = \sum_{i=1}^n\int_0^t
  \frac{S_2}{S_0}(s,\beta)-E(s,\beta)^{\otimes 2}\,dN_i(s) = \int_0^t V(s,\beta)\,dN_{.}(s),
\end{align}
and hence minus the derivative of the score is $I(\tau,\beta)$ (i.e., the information).
The \emph{estimated} martingales residual process is given by
\begin{align}
  \begin{split}
    \widehat{M}_i(t) &= N_i(t) - \widehat{\Lambda}_i(t) = N_i(t) -
    \int_0^t
    Y_i(s)\exp\left\{X_i^T(s)\widehat{\beta}\right\}\,d\widehat{\Lambda}_0(s) \\
    &= N_i(t) - \int_0^t
    Y_i(s)\exp\left\{X_i^T(s)\widehat{\beta}\right\}\frac{1}{S_0(s,\widehat{\beta})}\,dN_{.}(s),
  \end{split}
\end{align}
(with the martingale residuals defined by evaluation in $\tau$), and
the estimated score process
\begin{align}
  \begin{split}
    U(\widehat{\beta},t) &= \sum_{i=1}^n\int_0^t
    X_i(s)\,d\widehat{M}_i(s) \\
    &= \sum_{i=1}^n \int_0^t
    \left\{X_i(s)-E(s,\widehat{\beta})\right\}\,dN_i(s),
  \end{split}
\end{align}
where $X_i(s)-E(s,\widehat{\beta})$ are the \emph{Schoenfeld residuals}.

To assess the proportional hazards assumption we will calculate the
Kolmogorov-Smirnov and Cramer-von-Mises test statistics of the
different coordinates of the observed score process. As in the previous section
we can simulate realizations under the null (proportional hazards).
The key result is that $n^{-1/2}\widehat{U}(\widehat{\beta},t)$ is
asymptotically equivalent to
\begin{align}
  n^{-1/2}\sum_{i=1}^\infty\Big\{M_{1i}(t) - I(t,\widehat{\beta})I(\tau,\widehat{\beta})^{-1}M_{1i}(\tau)\Big\},
\end{align}
with 
\begin{align}
  M_{1i}(t) = \int_0^\tau \left\{X_i(s) - e(s,\beta_0)\right\}\,dM_i(s),
\end{align}
where $e(t,\beta_0) \overset{\pr}{=} \lim_{n\to\infty} E(t,\beta_0)$
(see \citep{MR2214443,lin_et_al_model_check_cox_biometrika_1993}), which follows from a Taylor expansion
around the true parameter $\beta_0$.  With the estimates plugged in we get
\begin{align}
  \begin{split}
    \widehat{M}_{1i}(t) &= \int_0^t
    (X_i(s)-E(s,\widehat{\beta}))\,dN_i(s) \\
    &\qquad - \int_0^t
    (X_i(s) -E(s,\widehat{\beta}))\frac{\exp(X_i^T(s)\beta)}{S_0(s,\widehat{\beta})}\,dN_{.}(s).
  \end{split}
\end{align}
Given observed times $(T_1,\ldots,T_n)$ and death-indicators
$\Delta_i$, ($X_i = X_i(T_i)$) we can implement this by
\begin{align}
  \begin{split}
    \one_{(T_i\leq
      t,\Delta=1)}\left\{X_i(s)-E(T_i,\widehat{\beta})\right\} &-
    \exp(X_i^T\beta)\widehat{\Lambda}_0(t) \\
    &\qquad +
    \exp(X_i^T\beta)\int_0^tS_1/S_0^2(s,\widehat{\beta})\,dN_{.}(s).
  \end{split}
\end{align}
Finally $n^{-1/2}M_{1i}(t)$ is asymptotically equivalent to $n^{-1/2}\sum_{i=1}^n
\widehat{M}_{1i}(t)G_i$ where the $G_i$'s are i.i.d. $\mathcal{N}(0,1)$.

\subsection{Software}

The described methods are implemented in the \proglang{R}-package
\pkg{gof} available from the Comprehensive R Archive Network \citep{R}.

The package has been designed to work directly on \code{lm},
\code{glm} and \code{coxph} objects \citep{survivalR}.  Additionally,
various aspects of latent variable models, fitted via the \pkg{lava}-package
\citep{lava}, can be diagnosed. 

The simulation routine is computational intensive and to obtain better
computing efficiency, the resampling routines was written in
\proglang{C++}. The implementation uses the \pkg{Scythe Statistical
  Library} \citep{Pemstein:Quinn:Martin:2007:JSSOBK:v42i12} which
among other things offers operator overloaded matrix operations making
the (linear) algebraic computations in the program close to
self-documenting.


\section{Examples}
In the following section the \pkg{gof} package will be demonstrated in
generalized linear models, a structural equation model and a Cox
regression model.

\subsection{Generalized linear models}
\label{sec:glm}
First we define a simple function that allows us to simulate data from
Binomial and Poisson regression models with link function $g$, and
covariates $X,Z\sim\mathcal{N}(0,1)$
\begin{align}
  g\left(\E[Y\mid X,Z]\right) = f(X,Z).
\end{align}
\begin{Schunk}
\begin{Sinput}
R> sim1 <- function(n,f=sum,family=binomial("logit")) {
   x <- rnorm(n)
   z <- rnorm(n)
   if (is.character(family)) family <- do.call(family,list())
   eta <- family$linkinv(apply(cbind(x,z),1,f))
   y <- switch(family$family,
               binomial= (eta>runif(n))*1,
               poisson= rpois(n,eta),
               eta)  
   return(data.frame(y,x,z))
 }
\end{Sinput}
\end{Schunk}

We first simulate binomially distributed observations and use a
complementary log-log link:
\begin{align} 
  \log\left(-\log\left[1-\E(Y\mid X,Z)\right]\right) = X+Z
\end{align}
\begin{Schunk}
\begin{Sinput}
R> d <- sim1(n=1000,family=binomial("cloglog"))
\end{Sinput}
\end{Schunk}
Next we fit both the correct model and the model with canonical link
\begin{Schunk}
\begin{Sinput}
R> l1 <- glm(y~x+z,d,family=binomial("cloglog"))
R> l2 <- glm(y~x+z,d,family=binomial("logit"))
\end{Sinput}
\end{Schunk}
Using the \code{cumres} method, we calculate the cumulative residual
process ordered by the predicted values and simulate 1,000 processes
from the null
\begin{Schunk}
\begin{Sinput}
R> library("gof")
\end{Sinput}
\end{Schunk}
\begin{Schunk}
\begin{Sinput}
R> (g1 <- cumres(l1,R=1000,variable="predicted"))
\end{Sinput}
\begin{Soutput}
Kolmogorov-Smirnov-test: p-value=0.466
Cramer von Mises-test: p-value=0.333
Based on 1000 realizations. Cumulated residuals ordered by predicted-variable.
---
\end{Soutput}
\end{Schunk}
\begin{Schunk}
\begin{Soutput}
Kolmogorov-Smirnov-test: p-value=0.466
Cramer von Mises-test: p-value=0.333
Based on 1000 realizations. Cumulated residuals ordered by predicted-variable.
---
\end{Soutput}
\end{Schunk}
\begin{Schunk}
\begin{Sinput}
R> (g2 <- cumres(l2,R=1000,variable="predicted"))
\end{Sinput}
\begin{Soutput}
Kolmogorov-Smirnov-test: p-value=0.014
Cramer von Mises-test: p-value=0
Based on 1000 realizations. Cumulated residuals ordered by predicted-variable.
---
\end{Soutput}
\end{Schunk}
\begin{Schunk}
\begin{Soutput}
Kolmogorov-Smirnov-test: p-value=0.014
Cramer von Mises-test: p-value=0
Based on 1000 realizations. Cumulated residuals ordered by predicted-variable.
---
\end{Soutput}
\end{Schunk}
There are clear indications, by both the supremum and CvM test, of
misspecification of the link function in model \code{l2}. To plot the
observed process and realizations from under the null (the number of
realization can be changed in the \code{cumres} call with the argument
\code{plots}), we can use the \code{plot} method
\begin{figure}[H]
  \centering
\begin{Schunk}
\begin{Sinput}
R> par(mfrow=c(1,2))
R> plot(g1,title="Model 'l1'"); plot(g2,title="Model 'l2'") 
\end{Sinput}
\end{Schunk}
\includegraphics{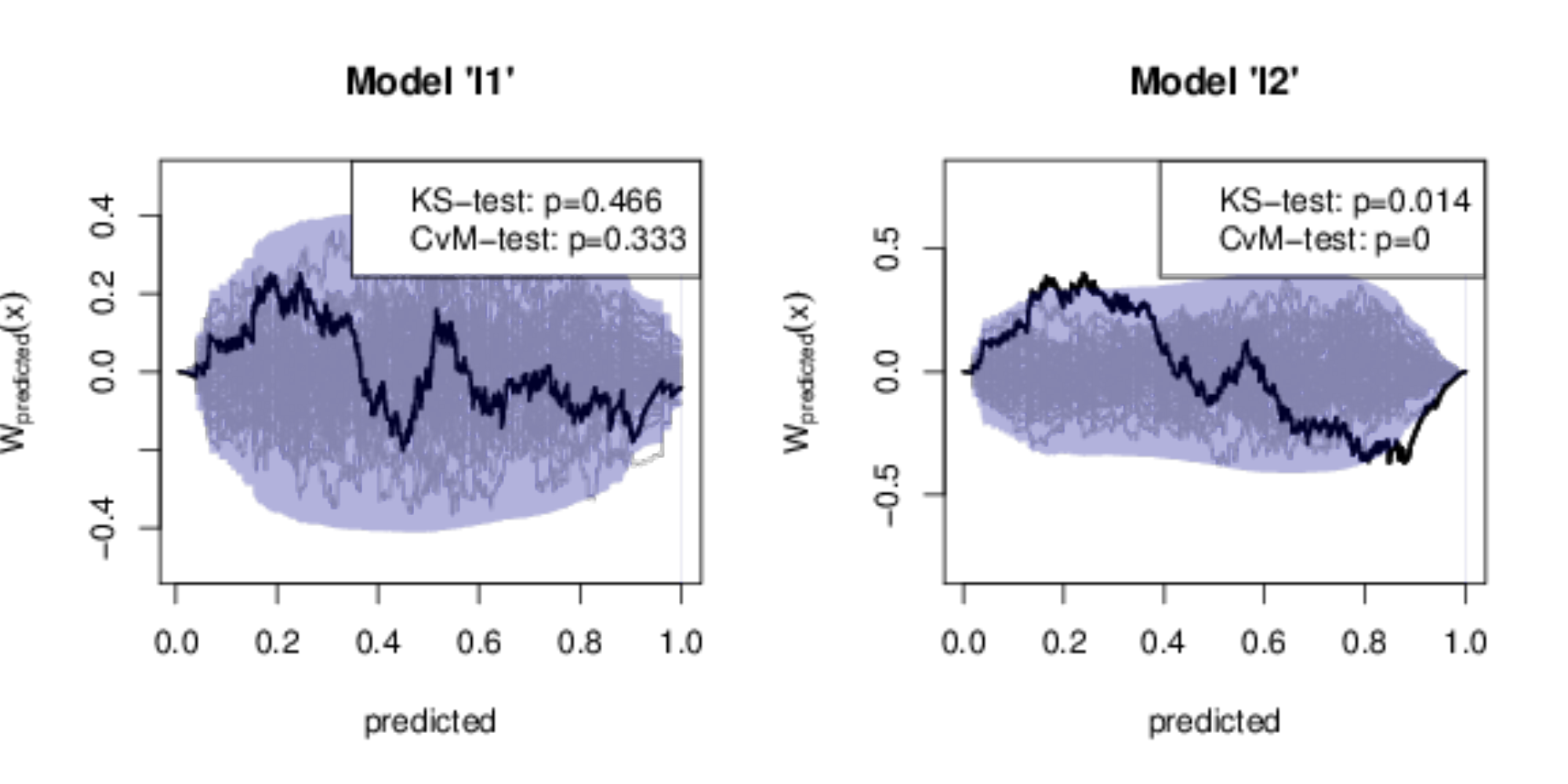}
\caption{Cumulative residual processes of model \code{l1} and
  \code{l2} with residuals ordered by the predicted response. The gray
  curves are 50 realizations from the null model.
  The transparent blue area defines a 95\% prediction band for all the
  simulated processes.
}
\end{figure}
It is evident from the plot (Figure~\ref{plot}), that the observed
process of model 2 is extreme.

\begin{figure}[H]
  \centering
\begin{Schunk}
\begin{Sinput}
R> par(mfrow=c(2,2))
R> plot(g,col="gray",col.ci="black",col.alpha=0.4,legend=NULL) 
\end{Sinput}
\end{Schunk}
\includegraphics{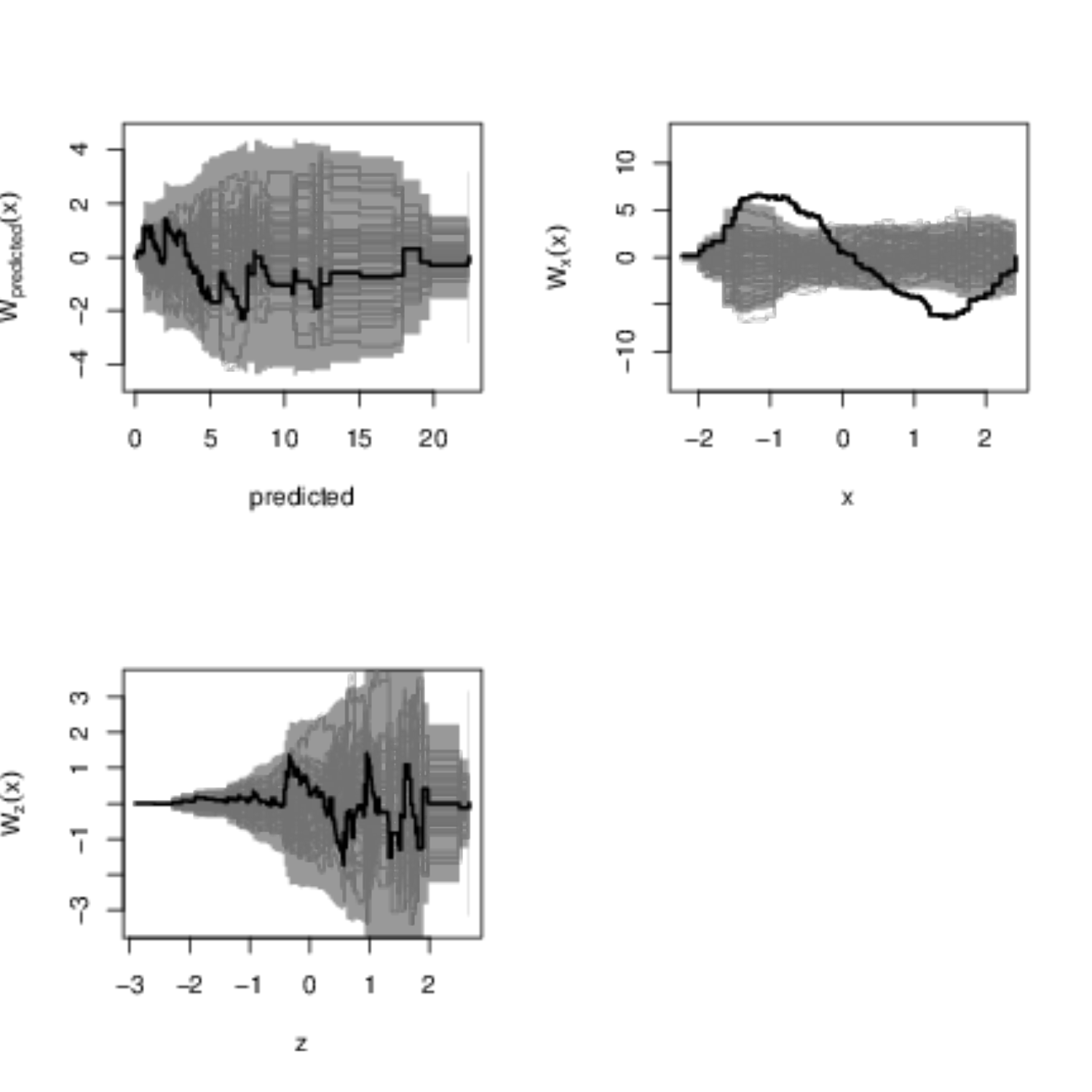}
\caption{Cumulative residual processes for the Poisson regression
  model \code{l}.\label{plot2}}
\end{figure}

Next we simulate data from a Poisson regression model
\begin{align}
  \log(\E(Y\mid X,Z)) = 0.5\cdot X^2+Z
\end{align}
\begin{Schunk}
\begin{Sinput}
R> d2 <- sim1(200,f=function(x) 0.5*x[1]^2+x[2],family=poisson())
\end{Sinput}
\end{Schunk}
and we fit a Poisson regression model but with misspecified functional
form of the covariate $X$
\begin{Schunk}
\begin{Sinput}
R> l <- glm(y~x+z,family=poisson(),data=d2)
\end{Sinput}
\end{Schunk}
Next we check the link function and functional form of both covariates 
\begin{Schunk}
\begin{Sinput}
R> (g <- cumres(l,R=2000)) 
\end{Sinput}
\begin{Soutput}
Kolmogorov-Smirnov-test: p-value=0.453
Cramer von Mises-test: p-value=0.547
Based on 2000 realizations. Cumulated residuals ordered by predicted-variable.
---
Kolmogorov-Smirnov-test: p-value=0.001
Cramer von Mises-test: p-value=0
Based on 2000 realizations. Cumulated residuals ordered by x-variable.
---
Kolmogorov-Smirnov-test: p-value=0.8185
Cramer von Mises-test: p-value=0.858
Based on 2000 realizations. Cumulated residuals ordered by z-variable.
---
\end{Soutput}
\end{Schunk}
\begin{Schunk}
\begin{Soutput}
Kolmogorov-Smirnov-test: p-value=0.453
Cramer von Mises-test: p-value=0.547
Based on 2000 realizations. Cumulated residuals ordered by predicted-variable.
---
Kolmogorov-Smirnov-test: p-value=0.001
Cramer von Mises-test: p-value=0
Based on 2000 realizations. Cumulated residuals ordered by x-variable.
---
Kolmogorov-Smirnov-test: p-value=0.8185
Cramer von Mises-test: p-value=0.858
Based on 2000 realizations. Cumulated residuals ordered by z-variable.
---
\end{Soutput}
\end{Schunk}
and we plot all processes (Figure~\ref{plot2}) while changing the color (and alpha
blending) of the realizations and prediction-band (setting \code{col}
or \code{col.ci} to NULL will disable either the realizations or the prediction-band)
Again, the misspecification (of the functional form of $X$) is evident from
the plots.

\subsection{Structural equation models}
\begin{figure}[h]
  \centering
\includegraphics{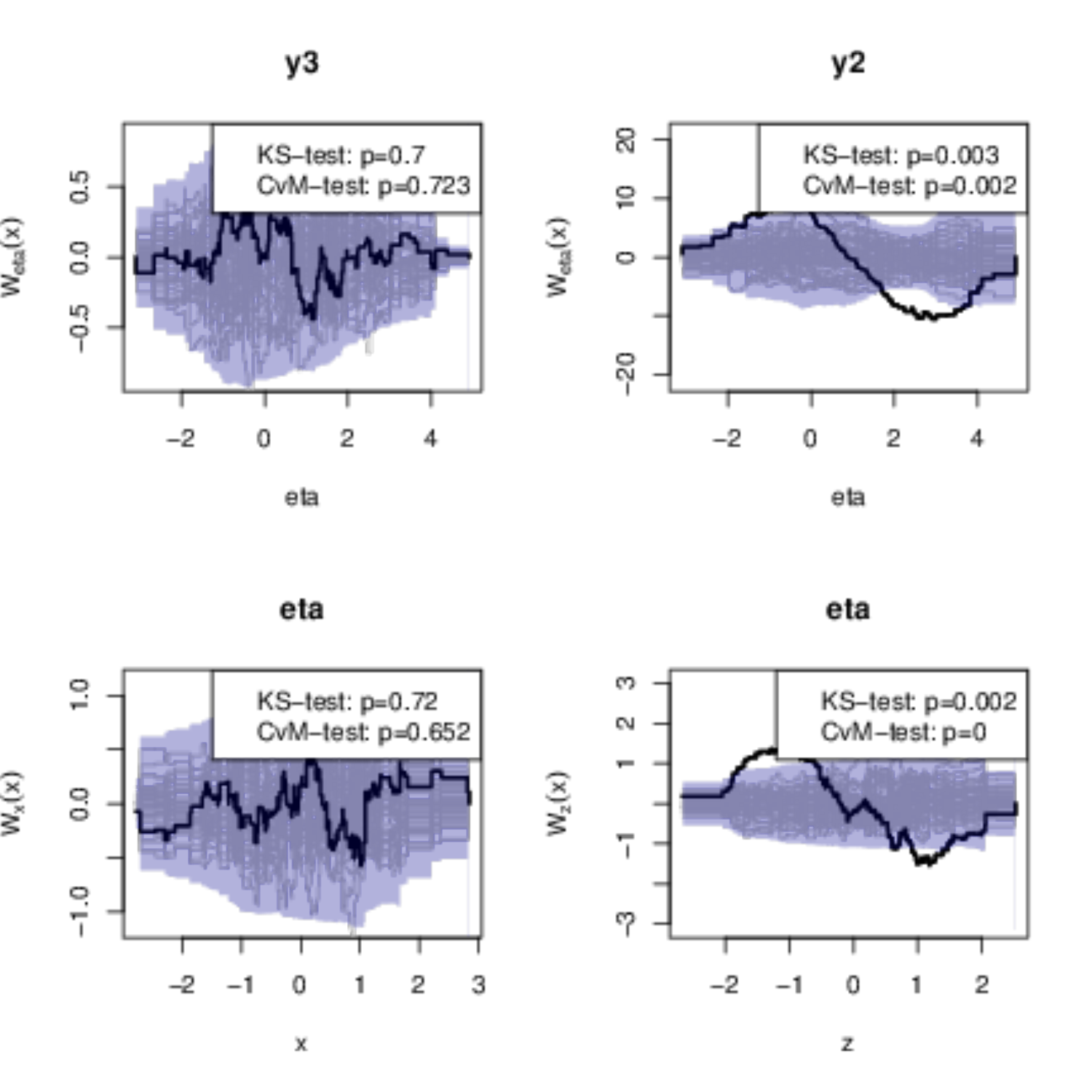}
\caption{\code{par(mfrow=c(2,2)); lapply(e.gof,plot)}. Selected
  cumulative residual processes for the structural equation model fit
  \code{e}. The top row shows the cumulative residuals of
  $\epsilon_{3i}$ and $\epsilon_{2i}$ (see (\ref{eq:measA})) ordered
  by $\E(\eta_i\mid X_i,Z_i)$.  The bottom row shows the cumulative
  processes of the predicted residual term, $\widehat{\zeta}_{i}$, of
  the latent variable ordered by each of the two covariates.  }\label{fig:csem}
\end{figure}

The \code{cumres} method is also available for structural equation
models fitted via the \code{lava} package \citep{lava}. As an example we will
examine a simple model, with three outcomes described by the equation
\begin{align}\label{eq:measA}
  Y_{ij} = \mu_j + \lambda_j\eta_i + \epsilon_{ij}, j=1,\ldots,3,
\end{align}
with $i=1,\ldots,n$ individuals and latent variable $\eta_i$. We also 
add a structural equation describing the latent variable
\begin{align}
  \eta_i = \beta_1\cdot X + \beta_2 Z + \zeta,
\end{align}
with covariates $X$ and $Z1$. The residual terms
$\epsilon_{i1},\ldots,\epsilon_{i3},\zeta$ are normally distributed and
independent. In \code{lava} we can specify the model as
\begin{Schunk}
\begin{Sinput}
R> library(lava)
R> m <- lvm(list(c(y1,y2,y3)~eta,eta~x+z))
R> latent(m) <- ~eta
\end{Sinput}
\end{Schunk}
We simulate 200 observations from a structural equation model like the
one defined above, with intercepts set to zero and all other parameters
equal to one, but with 
\begin{align}
  Y_{i2} = \eta^2 + \epsilon_{i2} \quad\text{and}\quad
  \eta_i = X + 0.5\cdot X^2 + Z + \zeta.
\end{align}
\begin{Schunk}
\begin{Sinput}
R> m0 <- m
R> functional(m0,y2~eta) <- function(x) x^2
R> functional(m0,eta~z) <- function(x) x+0.5*x^2
R> d <- sim(m0,200)
\end{Sinput}
\end{Schunk}
Next we find the MLE of the first model
\begin{Schunk}
\begin{Sinput}
R> (e <- estimate(m,d))
\end{Sinput}
\end{Schunk}
\begin{Schunk}
\begin{Soutput}
                    Estimate Std. Error  Z-value   P-value
Measurements:                                             
   y2<-eta           2.21830    0.24535  9.04147    <1e-12
   y3<-eta           0.99314    0.05380 18.45847    <1e-12
Regressions:                                              
   eta<-x            1.00280    0.09820 10.21176    <1e-12
   eta<-z            1.08055    0.09790 11.03729    <1e-12
Intercepts:                                               
   y2                2.84517    0.48757  5.83545 5.365e-09
   y3               -0.05982    0.09412 -0.63558    0.5251
   eta               0.50307    0.10623  4.73553 2.185e-06
Residual Variances:                                       
   y1                0.67836    0.15287  4.43739          
   y2               40.57804    4.22914  9.59488          
   y3                0.92828    0.16468  5.63682          
   eta               1.57358    0.21605  7.28351          
\end{Soutput}
\end{Schunk}
and as an example we cumulate the predicted residual terms of $Y_3$
and $Y_2$ against $\E(\eta_i\mid X_i)$, and the residual term of
$\eta_i$ against the two covariates. 
\begin{Schunk}
\begin{Sinput}
R> e.gof <- cumres(e,list(y3~eta,y2~eta,eta~x,eta~z),R=1000)
\end{Sinput}
\end{Schunk}
From the cumulative residual plots (see Figure~\ref{fig:csem}) we
clearly see the misspecification in the measurement model of the
second outcome (with the observed process also indicating a quadratic
form), and also the wrongly specified functional form of $X$.

For complete flexibility the \code{cumres} method can be used with the syntax
\code{cumres(model,y,x,...)}, where \code{y} is a function of the
model parameters returning the residuals of interest, and \code{x} can
be any vector to order the residuals by. Typically \code{y} will be
defined via the \code{predict} method of a \code{lvmfit} object
(a \code{lava} model object).
\clearpage

\subsection{Cox regression - Mayo clinic PBC data}
As an example of checking the proportional hazards assumption in a Cox
model, we will analyze the Mayo Clinic PBC data. 
\cite{dickson89:_progn} suggested a Cox model for analyzing the
survival of the liver disease patient with 5 covariates: age, edema
status, logarithmic serum bilirubin, logarithmic standardized blood
clotting time, and logarithmic serum albumin: 
\begin{Schunk}
\begin{Sinput}
R> library("survival")
R> data("pbc")
R> pbc.cox <- coxph(Surv(time,status==2)~age+edema+log(bili)+
 	log(protime)+log(albumin), data=pbc)
\end{Sinput}
\end{Schunk}

To check the proportional hazards assumption, we examine the score
process vs. follow-up time:
\begin{Schunk}
\begin{Sinput}
R> pbc.gof <- cumres(pbc.cox,R=2000)
\end{Sinput}
\end{Schunk}
and plot the observed process with realizations from the null
\begin{figure}[H]
  \centering
\begin{Schunk}
\begin{Sinput}
R> par(mfrow=c(2,3))
R> plot(pbc.gof,legend=FALSE)
\end{Sinput}
\end{Schunk}
\includegraphics{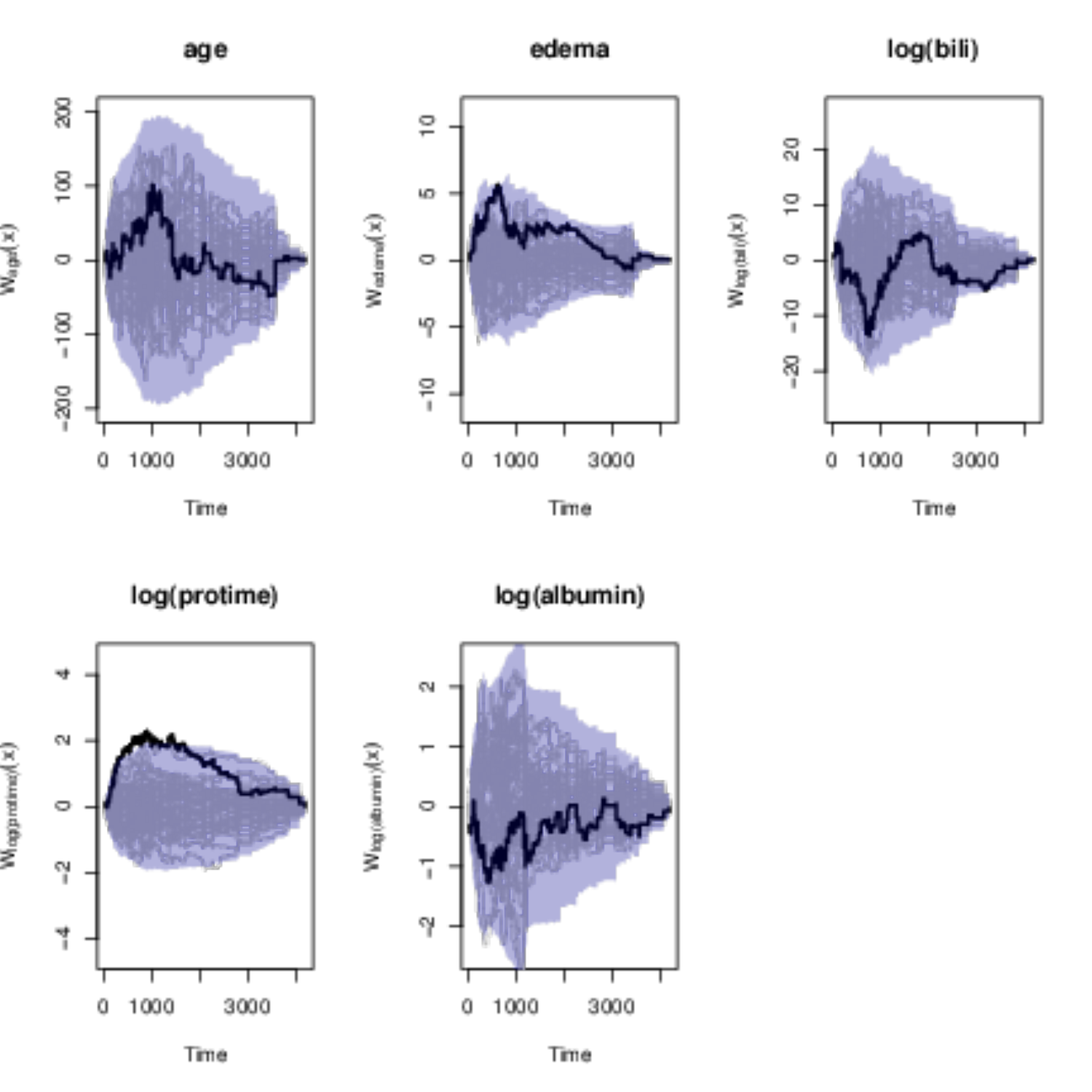}
\caption{Cumulative score processes for the Cox regression analysis of the PBC
  data, \code{pbc.cox}.}
\end{figure}
There are clear indication of violation of the proportional hazards
assumption for blood clotting time (protime), and indication of problems with
the edema variable.  To remedy the non-proportionality, time-varying
covariate effects could be introduced to the model, e.g.,
\begin{Schunk}
\begin{Sinput}
R> library("timereg")
R> pbc.caalen <- cox.aalen(Surv(time,status==2) ~ prop(age) + prop(edema) + 
 	   prop(bili) + protime, data=pbc, n.sim=500)
\end{Sinput}
\end{Schunk}



\section{Conclusion}
The package \pkg{gof} adds a valuable tool to the model diagnostics
toolbox and gives an objective method for evaluating the linearity
assumptions in the generalized linear model and linear structural
equation models. Extensions to other models such as the linear mixed
model can be implemented using the \proglang{C++} interface as used by
the \code{cumres} method for \code{glm} and \code{lvm} objects.

\section{Acknowledgments}
This work was supported by The Danish Agency for Science, Technology
and Innovation.

\bibliographystyle{elsarticle-harv}
\bibliography{gof}

\begin{thebibliography}{16}
\expandafter\ifx\csname natexlab\endcsname\relax\def\natexlab#1{#1}\fi
\expandafter\ifx\csname url\endcsname\relax
  \def\url#1{\texttt{#1}}\fi
\expandafter\ifx\csname urlprefix\endcsname\relax\def\urlprefix{URL }\fi

\bibitem[{Bollen(1989)}]{MR996025}
Bollen, K.~A., 1989. Structural equations with latent variables. Wiley Series
  in Probability and Mathematical Statistics: Applied Probability and
  Statistics. John Wiley \& Sons Inc., New York, a Wiley-Interscience
  Publication.

\bibitem[{Cox(1972)}]{cox72}
Cox, D.~R., 1972. Regression models and life tables. J. Roy. Stat. Soc. Ser. B
  34, 406--424.

\bibitem[{Dickson et~al.(1989)Dickson, Grambsch, Fleming, Fisher, and
  Langworthy}]{dickson89:_progn}
Dickson, E., Grambsch, P., Fleming, T., Fisher, L., Langworthy, A., 1989.
  Prognosis in primary biliary cirrhosis: model for decision making. Hepatology
  10, 1--7.

\bibitem[{Holst and Budtz-Joergensen(2012)}]{lava}
Holst, K.~K., Budtz-Joergensen, E., 2012. Linear latent variable models: The
  lava-package. Computational
  StatisticsHttp://dx.doi.org/10.1007/s00180-012-0344-y.

\bibitem[{Lin et~al.(1993)Lin, Wei, and
  Ying}]{lin_et_al_model_check_cox_biometrika_1993}
Lin, D.~Y., Wei, L.~J., Ying, Z., 1993. Checking the {C}ox model with
  cumulative sums of martingale-based residuals. Biometrika 80~(3), 557--572.

\bibitem[{Lin et~al.(2002)Lin, Wei, and
  Ying}]{lin02:_model_check_techn_based_cumul_resid}
Lin, D.~Y., Wei, L.~J., Ying, Z., 2002. {M}odel-{C}hecking {T}echniques {B}ased
  on {C}umulative {R}esiduals. Biometrics 58, 1--12.

\bibitem[{Martinussen and Scheike(2006)}]{MR2214443}
Martinussen, T., Scheike, T.~H., 2006. Dynamic regression models for survival
  data. Statistics for Biology and Health. Springer-Verlag, New York.

\bibitem[{McCullagh and Nelder(1983)}]{MR727836}
McCullagh, P., Nelder, J.~A., 1983. Generalized linear models. Monographs on
  Statistics and Applied Probability. Chapman \& Hall, London.

\bibitem[{Pan and Lin(2005)}]{ZhiyingPan2005}
Pan, Z., Lin, D.~Y., 2005. Goodness-of-fit methods for generalized linear mixed
  models. Biometrics 61, 1000--1009.

\bibitem[{Pemstein et~al.(2011)Pemstein, Quinn, and
  Martin}]{Pemstein:Quinn:Martin:2007:JSSOBK:v42i12}
Pemstein, D., Quinn, K.~M., Martin, A.~D., 6 2011. The scythe statistical
  library: An open source c++ library for statistical computation. Journal of
  Statistical Software 42~(12), 1--26.
\newline\urlprefix\url{http://www.jstatsoft.org/v42/i12}

\bibitem[{{R Core Team}(2012)}]{R}
{R Core Team}, 2012. R: A Language and Environment for Statistical Computing. R
  Foundation for Statistical Computing, Vienna, Austria, {ISBN} 3-900051-07-0.
\newline\urlprefix\url{http://www.R-project.org}

\bibitem[{S{\'a}nchez et~al.(2009)S{\'a}nchez, Houseman, and
  Ryan}]{BNSanchez2009}
S{\'a}nchez, B.~N., Houseman, E.~A., Ryan, L.~M., 2009. Residual-based
  diagnostics for structural equation models. Biometrics 65~(1), 104--115.

\bibitem[{Shorack and Wellner(1986)}]{shorack86:_empir_proces_applic_statis}
Shorack, G.~R., Wellner, J.~A., 1986. Empirical Processes with Applications to
  Statistics. John Wiley \& Sons, New York.

\bibitem[{Su and Wei(1991)}]{MR1137124}
Su, J.~Q., Wei, L.~J., 1991. A lack-of-fit test for the mean function in a
  generalized linear model. Journal of American Statistical Association
  86~(414), 420--426.

\bibitem[{Therneau and original {R} port~by Thomas~Lumley(2013)}]{survivalR}
Therneau, T., original {R} port~by Thomas~Lumley, 2013. survival: Survival
  analysis, including penalised likelihood. {R} package version 2.37-4.
\newline\urlprefix\url{http://CRAN.R-project.org/package=survival}

\bibitem[{Zeileis(2006)}]{Zeileis:2006nx}
Zeileis, A., 2006. Object-oriented computation of sandwich estimators. Journal
  of Statistical Software 16~(9), 1--16.
\newline\urlprefix\url{http://www.jstatsoft.org/v16/i09/.}

\end{thebibliography}

\end{document}